\documentclass[prc,twocolumn,showpacs,amsmath,amssymb]{revtex4}
\usepackage{graphicx}
\usepackage{enumerate}

\def \mate<#1|#2|#3>{\mbox{$\langle {#1}|\,{#2}\,|{#3}\rangle$}}

\begin{document}

\title{Medium-heavy nuclei from nucleon-nucleon interactions in lattice QCD}

\author{
Takashi Inoue$^{1}$,
Sinya Aoki$^{2,3}$,
Bruno Charron$^{4,5}$,
Takumi Doi$^{4}$,
Tetsuo Hatsuda$^{4,6}$,\\
Yoichi Ikeda$^{4}$,
Noriyoshi Ishii$^{7}$,
Keiko Murano$^{7}$,
Hidekatsu Nemura$^{3}$,
Kenji Sasaki$^{3}$\\ (HAL QCD Collaboration)\\
}

\affiliation{
$^1${Nihon University, College of Bioresource Sciences, Kanagawa 252-0880, Japan}\\
$^2${Yukawa Institute for Theoretical Physics, Kyoto University, Kyoto 606-8502, Japan }\\
$^3${Center for Computational Sciences, University of Tsukuba, Ibaraki 305-8571, Japan}\\
$^4${Theoretical Research Division, Nishina Center, RIKEN, Saitama 351-0198, Japan}\\
$^5${Department of Physics, The University of Tokyo, Tokyo 113-0033, Japan}\\
$^6${Kavli IPMU (WPI), The University of Tokyo, Chiba 277-8583, Japan}\\
$^7${Research Center for Nuclear Physics (RCNP), Osaka University, Osaka 567-0047, Japan}
}

\begin{abstract}
On the basis of the Brueckner-Hartree-Fock method with the nucleon-nucleon forces obtained from 
lattice QCD simulations, the properties of the medium-heavy doubly-magic nuclei such as $^{16}$O and
$^{40}$Ca are investigated. We found that those nuclei are bound for the pseudo-scalar meson
mass $M_{\rm PS}\simeq$ 470 MeV. The mass number dependence of the binding energies,
single-particle spectra and density distributions are qualitatively consistent with those expected
from empirical data at the physical point, although these hypothetical nuclei at heavy quark mass 
have smaller binding energies than the real nuclei.  
\end{abstract}
\maketitle

Studying the ground and excited states of finite nuclei and nuclear matter on the basis
of the quantum chromodynamics (QCD) has been one of the greatest challenges in modern nuclear physics.
Thanks to the recent advances in lattice QCD, we now have two major approaches to attack
this long-standing problem: The first approach is to simulate finite nuclei (systems with total baryon number $A$)
directly on the lattice~\cite{Yamazaki:2009ua,Beane:2011iw}. 
The second approach is to calculate the properties of finite nuclei and nuclear matter by using  
nuclear many-body techniques combined with the nuclear forces obtained from 
lattice QCD~\cite{Ishii:2006ec}. There is also a third
approach where nuclear many-body techniques are combined with the nuclear forces  from
chiral perturbation theory (see e.g. \cite{Gezerlis:2013ipa} and references therein); 
it has a close connection with the second approach
through the short distance part of the nuclear forces.

In this article, we will report a first exploratory attempt to study the structure of medium-heavy
nuclei ($^{16}$O and $^{40}$Ca) on the basis of the second approach by HAL QCD Collaboration~\cite{Ishii:2006ec}.  
Before going into the details, let us first summarize several limitations of the
first approach (direct QCD simulations of finite nuclei): 
(i) The number of quark contractions sharply increases for larger $A$, which makes the calculation prohibitively expensive.
Even with the help of newly discovered contraction algorithms \cite{Doi:2012xd},
it is still unrealistic to make simulations for medium-heavy nuclei with controlled $S/N$ on lattice.
(ii) The energy difference between the ground state and excited states, $\Delta E$, 
is about the QCD scale ($\sim$ 200 MeV) for single hadrons, while it becomes O(10)-O(100) times smaller for finite nuclei, 
which implies that extremely large Euclidean time $t \simeq 1/\Delta E \sim 100$ fm or more
is necessary to obtain sensible nuclear spectra;
(iii) The larger spatial lattice volume $V$ becomes necessary for larger nuclei.
This poses a challenge particularly for heavy nuclei and/or neutron-rich nuclei.
(iv) Analyzing the detailed spatial structure of nuclei (e.g. the 3$\alpha$ configuration
of the Hoyle state of $^{12}$C known to be crucial for the stellar nucleosynthesis) 
requires much more efforts beyond the calculation of binding energies.
     
The basic strategy of the second approach is to start with the lattice QCD simulations of nuclear forces 
in the form of the $A$-body potentials ($A=2,3,\cdots$). The nuclear structures can  then be 
calculated by the nuclear many-body techniques with the simulated potentials as inputs.
This two-step approach with  the ``potential" (the interaction kernel) 
as an intermediate tool provides not only a close link to the traditional nuclear physics 
but also a clue to overcoming the limitations (i)-(iv) mentioned above:
(i) The effect of the $A$-body potentials would decrease as $A$ increases for finite nuclei,
since the empirical saturation density $\rho_0$=0.16/fm$^3$ is rather low.
Then, we can focus mainly on the 2-body, 3-body and possibly 4-body potentials, 
exploiting the modern contraction algorithm~\cite{Doi:2012xd}.
(ii) Separation of the ground state and the excited states is not necessarily
to obtain the potentials as long as the system is below the pion production threshold~\cite{Ishii:2006ec}.
In other words, all of the information for $t > 1$ fm outside the range of inelastic region
can be  used to extract the potentials. 
(iii) The potentials among nucleons are always short ranged independent of $A$, so that they are
insensitive to the lattice volume~\cite{Inoue:2010es}.
(iv) Once the potentials in the continuum and infinite volume limit are obtained,
various observables can be obtained, e.g. the scattering
phase shifts, the nuclear binding energies, level structures, density distributions, etc.

As a first exploratory attempt, we limit ourselves to the two-body potentials in the $S$ and $D$ waves
in this article to study the structure of $^{16}$O and $^{40}$Ca. 
These potentials were previously obtained in ref.~\cite{Inoue:2011ai}
where the Nambu-Bethe-Salpeter (NBS) wave functions between two baryons simulated on the lattice 
are translated into the two-body potentials on the basis of the HAL QCD method (reviewed in the last reference of ~\cite{Ishii:2006ec}).
The resultant potentials in the nucleon-nucleon channel were applied to $^4$He with stochastic variational method 
in ref.~\cite{Inoue:2011ai} and to nuclear matter with Brueckner-Hartree-Fock (BHF) method in ref.~\cite{Inoue:2013nfe}.

We employ the standard BHF theory to calculate finite nuclei~\cite{RingSchuck}:
The main reason is that the BHF theory is simple but quantitative enough to grasp the essential part of physics,
so that it is a good starting point before making precise calculations using 
sophisticated {\it ab initio} methods such as
the Green's function Monte Carlo method~\cite{Pieper:2007ax},
no-core shell model~\cite{Navratil:2009ut,Shimizu:2012mv},
coupled-cluster theory~\cite{Hagen:2010gd},
unitary-model-operator approach~\cite{Fujii:2009bf},
self-consistent Green's function method~\cite{Dickhoff:2004xx},
and in-medium similarity renormalization group approach~\cite{Tsukiyama:2010rj}.

Let us briefly recapitulate the basic equations in the BHF theory for finite nuclei to set our notations.
The effective nucleon-nucleon interaction is dictated by the $G$ matrix satisfying the Bethe-Goldstone equation 
\begin{equation}
 G(\omega)_{ij,kl} \,=\, V_{ij,kl} \,+\,
 \frac12 \, \sum_{m,n}^{\mbox{\tiny un-occ}} \, \frac{V_{ij,mn}\,G(\omega)_{mn,kl}}{\omega - e_m - e_n + i\epsilon},
\label{eqn:gmat}
\end{equation}
where indices $i$ to $n$ stand for single-particle eigenstates,
$V$ is the bare $N\!N$ potential, and the sum is taken for un-occupied states.
Given $G$,the single-particle potential $U$ is written as
$ U_{ab} = \sum_{c,d} G(\tilde{\omega})_{ac,bd}~\rho_{dc}$,
where the indices $a,b,c,d$ are the labels for the harmonic-oscillator (HO)  basis.
The density matrix $\rho$  in this basis is given by
$ \rho_{ab} = \sum_{i}^{\rm occ}  \Psi^{i}_a \Psi^{i*}_b$,
where $\Psi^{i}$ is a solution of the Hartree-Fock equation,
\begin{equation}
 \left[ K + U \right] \Psi^{i} = e_i \Psi^{i}.
\label{eqn:scheq}
\end{equation}
with $K$ being the kinetic energy operator. After determining
$G$, $U$, $\rho$, $\Psi^{i}$, and $e_i$ self-consistently,
the ground state energy of a nucleus is obtained  as
\begin{equation}
 E_{0} = \sum_{a,b} \left[ K_{a b} + \frac12 U_{a b} \right] \rho_{b a} - K_{\rm cm}.
\label{eqn:gsene}
\end{equation}
Here $K_{\rm cm}$ corresponds to the subtraction of the spurious center-of-mass motion.

\begin{table}[t]
\caption{Masses of pseudo-scalar meson $M_{\rm PS}$, vector meson  $M_{\rm V}$ and octet baryon $M_{\rm B}$
in our calculation taken from ~\cite{Inoue:2011ai}. Statistical error is given in parentheses.}
\label{tbl:mass}
 \begin{tabular}{ccc}
   \hline  \hline
   ~~$M_{\rm PS}$ [MeV]~~ &  ~~$M_{\rm V}$ [MeV]~~ &  ~~$M_{\rm B}$ [MeV]~~ \\
   \hline 
      1170.9(7) &   1510.4(0.9) & 2274(2) \\
      1015.2(6) &   1360.6(1.1) & 2031(2) \\
    ~\,836.5(5) &   1188.9(0.9) & 1749(1) \\
    ~\,672.3(6) &   1027.6(1.0) & 1484(2) \\
    ~\,468.6(7) & ~\,829.2(1.5) & 1161(2) \\
   \hline  \hline
 \end{tabular}
\end{table}

\begin{figure}[t]
\includegraphics[width=0.4\textwidth]{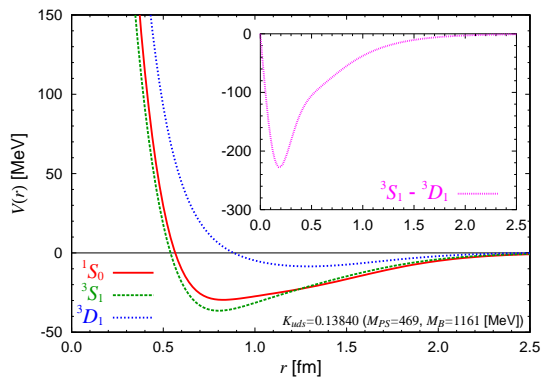}
\caption{Nucleon-nucleon potentials for S and D waves in lattice QCD at $M_{\rm PS}\simeq$ 470 MeV.
The lines are obtained by the least-chi-square fit to the lattice data.}
\label{fig:pot_K13840}
\end{figure} 

For the bare $N\!N$ potentials to be used in eq.(\ref{eqn:gmat}), we adopt those obtained on a (4 fm)$^3$ lattice with
five different quark masses in the flavor-$SU(3)$ limit~\cite{Inoue:2011ai} as summarized in Table~\ref{tbl:mass}.
As shown in Fig.~\ref{fig:pot_K13840}, the lattice $N\!N$ potentials in $S$ and $D$-waves at the pseudo-scalar meson
mass $M_{\rm PS} \simeq$ 470 MeV share common features with phenomenological potentials, i.e.,
a strong repulsive core at short distance, an attractive pocket at intermediate distance, and a strong $^3S_1$-$^3D_1$ coupling.
Although the potentials reproduce qualitative features of experimental phase-shifts,
the net attraction is still weak to form a deuteron bound state~\cite{Inoue:2011ai}, 
while it is strong enough to have saturation of symmetry nuclear matter (SNM) ~\cite{Inoue:2013nfe}. 

Using these lattice $N\!N$ potentials, together with the nucleon mass, as inputs,
we carry out the BHF calculation for the ground states of $^{16}$O and $^{40}$Ca nuclei.
We choose these nuclei since they are iso-symmetric, doubly magic, and spin saturated,
and hence we can assume spherically symmetric nucleon distribution.
Due to the limitation of  available  lattice $N\!N$ potentials at present,
we include 2-body $N\!N$ potentials only in $^1S_0$, $^3S_1$ and $^3D_1$ channels.
The Coulomb force between protons is not taken into account for simplicity.
We follow refs.~\cite{Daveis,Sauer} about the numerical procedure of BHF calculation, i.e.,
we solve eq.(\ref{eqn:gmat}) by separating the relative and center-of-mass coordinates
using the Talmi-Moshinsky coefficient, and adopt the so-called $Q/(\omega - QKQ)Q$ choice,
where $Q$ is the Pauli exclusion operator
for which we use a harmonic-oscillator one at first then use a self-consistent one for the last few iterations.
In eq.(\ref{eqn:gsene}), the center of mass correction is estimated as 
$K_{\rm cm} \simeq \frac{3}{4}\hbar \omega$
with $\omega$ being the a HO frequency which reproduces the root-mean-square (RMS) radius
of the matter distribution obtained by the BHF calculation.

\begin{figure}[t]
\includegraphics[width=0.4\textwidth]{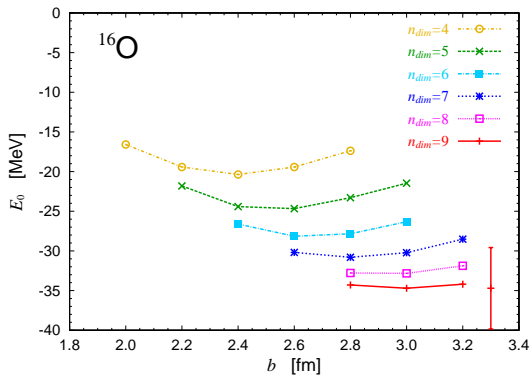}
\caption{Ground state energy of $^{16}$O at $M_{\rm PS}\simeq$ 470 MeV as a function of $b$ at several $n_{\rm dim}$.}
\label{fig:16Odepend}
\end{figure} 

Figure~\ref{fig:16Odepend} shows the ground state energy of $^{16}$O at $M_{\rm PS}\simeq$ 470 MeV,
as a function of the width parameter $b$ of the HO wave function with increasing number of HO basis $n_{\rm dim}$.
The solid vertical bar at the rightmost point represents the error for $E_0$ of about $\pm$10\%
at $b=3$ fm and $n_{\rm dim}=9$. It originates from the statistical error of our lattice QCD simulations
estimated by the Jackknife analysis with the bin-size of 360 for 720 measurements as was done in ref.~\cite{Inoue:2013nfe}.
Almost the same errors apply to other $E_0$ in the figure.
A similar figure for $^{40}$Ca is obtained for the same quark mass.
As $n_{\rm dim}$ increases, the binding energy $|E_0|$ increases with the optimal $b$ shifting to larger values.  
From these results, we can definitely say that self-bound systems are formed in both nuclei at this lightest quark mass,
corresponding to $M_{\rm PS}\simeq 470$ MeV and $M_{\rm B}\simeq 1160$ MeV. 
On the other hand, the existence of deeply bound nuclei is excluded for the other four heavier quark masses,
since we do not find $E_0<0$. 

\begin{figure}[t]
\includegraphics[width=0.4\textwidth]{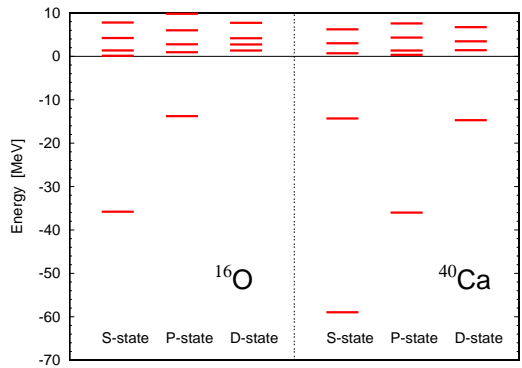}
\caption{Single particle levels of $^{16}$O and  $^{40}$Ca nuclei at $M_{\rm PS}\simeq$ 470 MeV.
Positive energy continuum states appear as discrete levels due to the finite number of bases.}
\label{fig:levels}
\end{figure}

In Figure~\ref{fig:levels}, single particle levels of $^{16}$O and $^{40}$Ca at $M_{\rm PS}\simeq$ 470 MeV,
are shown for the optimal width parameter with the largest HO basis;  $b=3.0$ fm and $n_{\rm dim}=9$.
In spite of the unphysical quark mass in our lattice QCD simulations, 
the obtained single particle levels have the similar magnitude expected for those nuclei in the real world.
Also, in the bound region, the level structure follows almost exactly the harmonic oscillator spectra
with $\hbar \omega \simeq 22-23$ MeV. 
Since the spin-orbit force is not included in our lattice nuclear force,
the spin-orbit splittings in the $P$ and $D$ states are not seen in the figure.

\begin{table}[t]
\caption{Single particle levels, total energy, and rms radius of $^{16}$O and $^{40}$Ca
at $M_{\rm PS} \simeq$ 470 MeV. Energies (radii) are in unit of MeV (fm).}
\label{tbl:structure}
 \begin{tabular}{c|cccc|cc|c}
   \hline \hline
            & \multicolumn{4}{c|}{Single particle level}
            & \multicolumn{2}{|c|}{Total energy} & \,Radius\, \\
            & ${1S}$  & ${1P}$  & ${2S}$  & ${1D}$  & $E_0$  & $E_{0}/A$ & $\sqrt{\langle r^2 \rangle}$  \\
\hline 
 $^{16}$O~  & \,$-35.8$\,  & \,$-13.8$\,  &              &           &   ~$-34.7$    & $-2.17$     &   $2.35$  \\
 $^{40}$Ca  & \,$-59.0$\,  & \,$-36.0$\,  & \,$-14.7$\,  & \,$-14.3$ & \,$-112.7$\,  & \,$-2.82$\, & \,$2.78$\,\\
   \hline \hline
 \end{tabular}
\end{table}

Table~\ref{tbl:structure} shows the single particle energies, total binding energies,
and rms radii of the matter distributions of $^{16}$O and $^{40}$Ca at $M_{\rm PS}\simeq$ 470 MeV
for $b=3.0$ fm and $n_{\rm dim}=9$.
Breakdowns of the total binding energies are 
\begin{eqnarray}
 ^{16}\mbox{O}:  &&\! E_0 = (259.6 - 10.3)  -284.0 = \,-34.7 \ {\rm MeV}, \qquad \\
 ^{40}\mbox{Ca}: &&\! E_0 = (813.4 - \,9.8) -916.3 =  -112.7 \ {\rm MeV}, \qquad
\label{eqn:break}
\end{eqnarray}
where the first, second, and third numbers are the kinetic energy, 
the center-of-mass correction and the potential energy, respectively. 
The total binding energy is obtained as a result of a large cancellation between kinetic energy and potential energy.
Principally due to the heavier quark mass in our calculation,
the obtained binding energies, $|E_0|$, are smaller than the experimental data,
$127.6$ MeV for $^{16}$O and $342.0$ MeV for $^{40}$Ca~\cite{Audi:1993zb}.

\begin{figure}[t]
\includegraphics[width=0.4\textwidth]{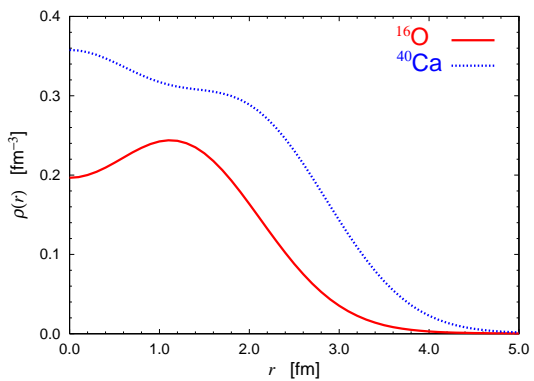}
\caption{Nucleon number density inside $^{16}$O and $^{40}$Ca at $M_{\rm PS}\simeq$ 470 MeV
as a function of distance from the center of the nucleus.}
\label{fig:density}
\end{figure} 

The rms radii of the matter distribution given in Table~\ref{tbl:structure} are calculated without the  
nucleon form-factor and  the center-of-mass correction.
We found that these radii are more or less similar to experimental charge radii 
(2.73 fm for $^{16}$O and 3.48 fm for $^{40}$Ca), although our quark mass is  heavier.
This is presumably due to  a cancellation between heavier nucleons and weaker nuclear forces than in the real world. 
Shown in Fig.~\ref{fig:density} is the spatial distribution of baryon number density $\rho(r)$ 
for $^{16}$O and $^{40}$Ca as a function of the  distance from the center of the nucleus.
The bump and dent at small distance originate from the shell structure
which are known to exist in the nuclear charge distribution extracted from the electron-nucleus  scattering experiments.
We also find that the central baryon density is as high as $2 \rho_0$ for $^{40}$Ca.
This is consistent with the fact that the saturation density of SNM for the present quark mass with 2-body $N\!N$ forces
is about $2.5 \rho_0$ \cite{Inoue:2013nfe}.

\begin{figure}[t]
\includegraphics[width=0.4\textwidth]{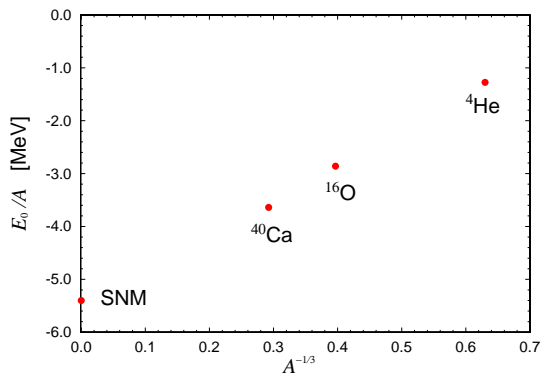}
\caption{Mass number $A$ dependence of nuclear energy per nucleon 
$E_0/A$ for $M_{\rm PS}\simeq$ 470 MeV.
The Bethe-Weizs{\"a}cker mass formula up to the second term, 
$E_0/A=-a_{\rm V}  - a_{\rm S} A^{-1/3}$,
corresponds to a straight line in this figure.}
\label{fig:adep}
\end{figure}

Finally, in Fig.~\ref{fig:adep}, the binding energies per particle $E_0/A$ for $A=4, 16, 40$, and $\infty$
obtained by using the same lattice potential at $M_{\rm PS}\simeq$ 470 MeV
are plotted  as a function of $A^{-1/3}$.
The stochastic variational method is used for $^4$He~\cite{Inoue:2011ai},
while the BHF method is used for SNM~\cite{Inoue:2013nfe}.
To make a fair comparison to these cases,
we carry out a linear extrapolation of the binding energies of 
${\rm ^{16}O}$ and ${\rm ^{40}Ca}$ to $n_{\rm dim}=\infty$
through the formula, $E_0(A;n_{\rm dim})= E_0(A;\infty) + c(A) /n_{\rm dim}$.
The linear formula fits our results well, although the convergence to $n_{\rm dim}=\infty$ is relatively slow.
(The faster convergence may be achieved by employing the approaches such as $V_{{\rm low} k }$ and the 
similarity renormalization group \cite{Hagen:2010gd}).
Our procedure leads to  $E_0(16;\infty)/16=-2.86$ MeV and  $E_0(40;\infty)/40=-3.64$ MeV. 
Note that these numbers are subject to the $\pm$10\% uncertainty due to the statistical
error in the $N\!N$ interactions from lattice QCD as mentioned already.
Although the magnitude of $|E_0/A|$ for $^{16}$O, $^{40}$Ca, and SNM are a factor of 3--4 smaller than the 
empirical values, its $A$ dependence is uniform and can be approximated by the
Bethe-Weizs{\"a}cker type mass formula, $E_0(A) = - a_{\rm V} A - a_{\rm S} A^{2/3}$, with 
$a_{\rm V}=5.46$ MeV and $a_{\rm S}=-6.56$ MeV. It would be interesting in the future to study
the quark mass dependences of $a_{\rm V,S}$ in the lighter quark mass region and investigate how
these coefficients approach the empirical values, $a_{\rm V}^{\rm phys}=15.7$ MeV 
and $a_{\rm S}^{\rm phys}=-18.6$ MeV.
  
In this Rapid Communication, we have shown that properties of medium-heavy nuclei can be deduced 
by combining the nuclear many-body method with the nuclear force obtained from lattice QCD simulations.
Using the BHF theory with 2-body $N\!N$ potentials  at $M_{\rm PS}\simeq$ 470 MeV,
we found bound nuclei for $^{16}$O and $^{40}$Ca, and we could extract
their binding energies, single-particle spectra, and density distributions.
Even though our setup is still primitive in various places, our results
demonstrate that the HAL QCD approach to nuclear physics is quite promising
for unraveling the structure of finite nuclei and infinite nuclear matter in a unified manner from QCD. 

In the present  study, we have neglected the nuclear forces in $P$, $F$ and higher partial-waves,
in particular the effect of the spin-orbit ($LS$) force: For nuclei with $A>40$,
the $LS$ force plays a crucial role in developing the  magic numbers. Therefore it will be an important next step
to include the $LS$ force recently extracted from lattice QCD simulations~\cite{Murano:2013xxa}. 
The 3-body force may also play an essential role for accurate determinations of the
binding energy and the structure of finite nuclei as well as nuclear matter.
Study of the three-nucleon force in QCD is also in progress~\cite{Doi:2011gq}. 
Finally, the masses of up and down quarks in this study are much heavier than the physical values.
We are currently working on the almost physical point lattice QCD simulations with
the lattice volume (8 fm)$^3$ on the K-computer at RIKEN AICS.
Lattice QCD potentials obtained in such simulations together with advanced 
nuclear many-body methods will open a new connection between QCD and nuclear physics.

\smallskip
\begin{acknowledgments}
This research is supported in part by Grant-in-Aid of MEXT-Japan 
for Scientific Research (B) 25287046, 24740146, (C) 26400281, 23540321
and SPIRE (Strategic Program for Innovative REsearch).
T.H. was partially supported by RIKEN iTHES Project.
\end{acknowledgments}


\end{document}